\documentclass[journal,draftcls,onecolumn,12pt,twoside]{IEEEtranTCOM}
\normalsize

\usepackage{cite} 
\usepackage{url}  
\usepackage{ifthen}  
\usepackage{multicol}   
\usepackage[utf8]{inputenc} 
\usepackage[dvips]{graphicx}
\usepackage{psfrag}
\usepackage{subfigure}
\usepackage{theorem}

\usepackage{enumerate}
\usepackage{times}
\usepackage{amsmath}
\usepackage{bm}
\usepackage{comment}
\usepackage{amsfonts}
\usepackage{multirow}

\usepackage{subfigure}

\DeclareMathOperator*{\argmax}{arg\,max}

\ifCLASSINFOpdf
\else
\fi

\begin{document}
%
\title{Performance Analysis of a Heterogeneous Traffic Scheduler using Large Deviation Principle}
%
%
%

\author{Rukhsana~Ruby,~David~G.~Michelson,~and~Victor~C.M.~Leung\thanks{R. Ruby, Prof. D. Michelson, and Prof. V. Leung are with the department of Electrical \& Computer Engineering at the University of British Columbia, Canada (email: rukhsana@ece.ubc.ca, davem@ece.ubc.ca, and vleung@ece.ubc.ca)}}
        

%
%

\markboth{Computer Networks}%
{Submitted paper}
%



\maketitle

\begin{abstract}
In this paper, we study the stability of light traffic achieved by a scheduling algorithm which is suitable for heterogeneous traffic networks. Since analyzing a scheduling algorithm is intractable using the conventional mathematical tool, our goal is to minimize the largest queue-overflow probability achieved by the algorithm. In the large deviation setting, this problem is equivalent to maximizing the asymptotic decay rate of the largest queue-overflow probability. We first derive an upper bound on the decay rate of the queueoverflow probability as the queue overflow threshold approaches infinity. Then, we study several structural properties of the minimum-cost-path to overflow of the queue with the largest length, which is basically equivalent to the decay rate of the largest queue-overflow probability. Given these properties, we prove that the queue with the largest length follows a sample path with linear increment. For certain parameter value, the scheduling algorithm is asymptotically optimal in reducing the largest queue length. Through numerical results, we have shown the large deviation properties of the queue length typically used in practice while varying one parameter of the algorithm.
\end{abstract}


%
\IEEEpeerreviewmaketitle

\section{Introduction}

Since wireless channels are time varying, by properly choosing scheduling algorithm, it is possible to achieve multi-user diversity, which enhances the performance of a system considerably. Most studies of scheduling algorithms have focused on optimizing the long term average throughput of users, i.e., stability. From the stability point of view, it is important to design an algorithm which schedules the transmissions in such a way that the queues are stabilized at given offered loads. Throughput optimal algorithms are considered as stable. Definition of a throughput optimal algorithm is as follows: at any offered load if any other algorithm can stabilize a system, the designated algorithm can stabilize the system as well. For example, MaxWeight (MW) and Exponential (EXP) algorithms are throughput optimal since they ensure stochastic stability of queues as long as such is feasible.

In other way, stability implies that average packet delay of users cannot reach infinity. Although stability is an important metric for designing a throughput optimal algorithm, for many delay-sensitive and guaranteed-rate traffic, it is not sufficient. For real-time applications such as voice and video, we often need to ensure a stronger condition that the packet delay should be upper bounded or the long-term throughput should be lower bounded by some threshold with high probability. One approach to quantify the requirements of these delay-sensitive and guaranteed-throughput applications is to enforce constraints on the probability of the queue overflow. In other words, we need to guarantee the smallest value of the probability that the largest queue length exceeds a given threshold, i.e., $P\left[\max_{1{\le}i{\le}N}Q_i(T) \ge B\right]$. Here, $N$ is the number of users, $Q_i(T)$ is the queue length of user $i$ at time $T$, and $B$ is the overflow threshold.


In this paper, at low traffic load, we study the stability property of a scheduling algorithm~\cite{RRuby15} usually used for the scheduling purpose in heterogeneous traffic networks. The system we consider is the downlink of a single cell in orthogonal frequency division multiplexing (OFDM)-based systems. There are many users in the network which are served by the base station. For the sake of simplicity, we consider, there is only one transmission which serves one user at a time. Perfect channel information is assumed to be available at the base station.

Since calculating the exact queue distribution is often mathematically intractable, for this problem, we use the large deviation theory. According to the property of large deviation, we reformulate the quality-of-service (QoS) constraint of delay-sensitive and guaranteed-throughput traffic in terms of the asymptotic decay rate of the queue-overflow probability as $B$ approaches infinity. Particularly, for this problem, large deviation theory is applicable since queue-overflow probability in a system with low load is very small.

Previously, large deviation theory was successfully applied to wireline networks as well as for channel state aware wireless scheduling algorithms. However, when applied to queue length aware wireless scheduling algorithms, this approach encounters a significant amount of technical difficulty. When the large deviation theory is applied to queue length aware scheduling algorithms, sample path large deviation is used while formulating the problem as a multidimensional calculus-of-variation (CoV) problem for finding the most likely path to overflow. However, for queue length based scheduling algorithms, this multidimensional CoV problem is very difficult to solve. 

In the literature, only some restricted cases have been solved: either restricted problem structures are assumed (e.g., symmetric users and ON-OFF channels~\cite{LeiYing06}), or the size of the system is very small (e.g., only two users)~\cite{ShakkottaiS08}. \cite{BilalSadiq092, BilalSadiq11} have used similar method in order to analyze MW scheduling rule for a system like ours. However, to make the problem simpler, the authors in this work used Lyapunov function to map multidimensional CoV problem to a one dimensional one. The result of this work is that MW algorithm maximizes the decay rate of the largest queue overflow probability when its one parameter approaches infinity. In an earlier work~\cite{Stolyar08}, the author showed that the EXP rule can maximize the decay rate of the largest queue overflow probability over all scheduling policies, however through the refined sample path large deviation principle (LDP) which is more technically involved. Since the algorithm studied in this paper and EXP scheduler both are queue length scale-variant, we have adopted the similar analytical technique as in~\cite{Stolyar08}. Analysis for the LOG rule scheduling algorithm has appeared in~\cite{Venkataramanan10}. Capturing a simplified network in the classical Markov Decision Process, they computed a mean delay optimal scheduling policy which exhibits radial sum rate monotonicity (RSM). The authors in this work also proved that LOG rule has RSM property which is absent in MW or EXP rule.


In this paper, we have shown a number of results for the minimum cost path to queue overflow event achieved by our scheduling scheme~\cite{RRuby15}. Presuming that the large deviation property exists for the largest queue, we have shown many properties for the sample path of the largest queue. We prove that the largest queue follows a linear increment before reaching a threshold length. Finally, we prove that our scheduling algorithm is asymptotically optimal in reducing the largest queue length over other scheduling algorithms for certain value of its parameter. Extensive simulation results verify the large deviation properties and optimality of our scheduling algorithm comparing with other algorithms in the literature.

The rest of the paper is organized as follows. We describe the system model and some preliminary definitions related to this problem in Section~\ref{sec:sysmodel} and Section~\ref{sec:preface}, respectively. We derive the generalized upper and lower bounds of the decay rate for the overflow probability of the largest queue length in Section~\ref{sec:up-low-bound}. Consequently, in the large deviation setting, we outline the properties of the sample path for the largest queue length in Section~\ref{sec:prop}. Our analytical study is justified with the simulation results in Section~\ref{sec:sim-res}, and Section~\ref{sec:concl} concludes this paper.

\section{System Model and Problem Formulation}
\label{sec:sysmodel}

We consider the downlink of a single cell in which the base station serves $N$ users. Time is considered as slotted and the channel state at each time slot is independent and identically distributed (i.i.d.) from one of $M$ possible states. Let $C(t)$ denote the state of the channel at time $t$, and the probability of $C(t)$ being $m$ is given by $p_m = \mbox{Pr}[C(t) = m], m=1,2,\cdots,M$. Let $\textbf{p} = [p_1,\cdots,p_M]$. As mentioned previously, the base station can serve one user at a time. Furthermore, at channel state $m$, if user $i$ is picked for service, its service rate is denoted by $F_m^i$.

The data packets for user $i$ arrive at a constant rate $\lambda_i$. Let $\bm{\lambda} = \left[\lambda_1,\cdots,\lambda_N\right]$. If $Q_i(t)$ denotes the backlog of user $i$ at time $t$, let $\textbf{Q}(t) = \left[Q_1(t),\cdots,Q_N(t)\right]$. In general, at time instant $t$, the decision of user selection is dependent on the global backlog $\textbf{Q}(t)$ and the channel state $C(t)$. Given this system, the behavior of the scheduling algorithm~\cite{RRuby15} is as follows. When the backlog of the users is $\textbf{Q}(t)$ and the state of the channel $C(t)$ is $m$, the algorithm chooses to serve user $i^*$ while obeying the following formula

\begin{equation}
\label{eq:eqn1}
i^* = \displaystyle\argmax_i1 - e^{\left[\rho_1 - \frac{F_m^i}{\max_{i{\in}\textbf{N}}F_m^i} + \rho_2 - \frac{Q_i(t)}{Q^{th}}\right]},
\end{equation} 

\noindent
where $Q^{th}$ is the acceptable queue level that ensures the QoS of all users. $\rho_1$ and $\rho_2$ are the parameters, the values of which are any natural number in between $0$ and $1$. If the tie happens, it is broken arbitrarily by choosing any tied user. The user selection rule of the above scheduling scheme is equivalent to 

\begin{equation}
i^* = \displaystyle\argmax_{i}- e^{\left[\rho_1 - \frac{F_m^i}{\max_{i{\in}\textbf{N}}F_m^i} + \rho_2 - \frac{Q_i(t)}{Q^{th}}\right]}.
\end{equation}  

Given the scheduling algorithm above, the system is presumed to be stable for the given offered load $\bm{\lambda}$. Specifically, we assume that $\bm{\lambda}(1 + \epsilon)$ is in the capacity region of the system. This implies that there exists $\left[{\phi}_m^i\right] \ge 0$ such that $\sum_{i=1}^N{\phi}_m^i=1$, $\forall{m{\in}[1,\cdots,M]}$, and

\[
{\lambda}_i \le \sum_{m=1}^Mp_m{\phi}_m^iF_m^i, \forall{i{\in}[1,\cdots,N]}.
\]

As the objective of this work is to prove that the aforementioned scheduler ensures the queue stability, we are interested in the probability that the largest backlog exceeds a certain threshold $B$ after certain time $T$, i.e., 

\begin{equation}
\label{eq:eqn2}
\mbox{Pr}\left[\displaystyle\max_iQ_i(T) \ge B\right].
\end{equation}

The probability in~(\ref{eq:eqn2}) is equivalent to the delay violation probability for delay-sensitive traffic or the throughput violation probability for guaranteed-throughput traffic. This is because these events are related by the relations $\mbox{Pr}\left[\mbox{Delay~for~user}~i \ge d_i\right] = \mbox{Pr}\left[Q_i(T) \ge \lambda_id_i\right]$ and $\mbox{Pr}\left[\mbox{Throughput~for~user}~i \le \zeta_i \right] = \mbox{Pr}\left[Q_i(T) \ge \lambda_i - \lambda_i\zeta_i\right]$.

Since the scheduling rule in (\ref{eq:eqn1}) is the function of $\textbf{Q}(t)$, calculating the probability in~(\ref{eq:eqn2}) is mathematically intractable. To circumvent this difficulty, we plan to use large deviation theory to minimize this probability. In the large deviation setting, minimizing the probability in (\ref{eq:eqn2}) is equivalent to maximizing the decay rate of this probability. Let define the upper and lower bounds of the decay rate for the probability in~(\ref{eq:eqn2}) as follows.

\begin{eqnarray}
\label{eq:eqn3}
I^{UP} = \displaystyle\liminf_{B \to \infty}\frac{1}{B}\mbox{log}P\left[\max_{1 \le i \le N}Q_i(T) \ge B\right] \\
I^{LOW} = \displaystyle\limsup_{B \to \infty}\frac{1}{B}\mbox{log}P\left[\max_{1 \le i \le N}Q_i(T) \ge B\right].
\end{eqnarray}


In the following sections, we will show that no scheduling algorithm can have a decay rate larger than a certain value $I_{opt}$, i.e., $I^{UP} \le I_{opt}$. Then, we will show that our scheduling algorithm asymptotically attains the decay rate $I_{opt}$ for certain value of its parameter ($Q^{th}$) and several properties for the sample path of the largest queue backlog.

\section{Preface}
\label{sec:preface}

It is already mentioned that our scheduling algorithm is not queue length scaling invariant. That implies, given the arrival rate, if we scale the channel rate process in some magnitude, the length of the resultant queues do not scale in the same magnitude. Hence, to study the properties of the scaled queue length, we need to understand the system process empirically. For this, we need to introduce additional functions associated with the system evolution. For $T \ge 0$, let

\[
F_i(T) = \sum_{t=1}^TA_i(t)~\mbox{and}~{\hat{F}}_i(T) = \sum_{t=1}^TD_i(t),
\] 

\noindent
where $F_i(T)$ and ${\hat{F}}_i(T)$ are the number of packets that arrive and depart for user $i$ over the time interval $\left[0,T\right]$. Furthermore, denote by $G_m(T)$ the total number of time slots when the channel is in state $m$ over the interval $\left[0,T\right]$, and by ${\hat{G}}_m^i(T)$ the number of time slots when the channel is in state $m$ and user $i$ is chosen for service. Note that $A_i(t)$ and $D_i(t)$ are the number of packets that arrive and depart for user $i$ at time slot $t$. Given these functions to evolve the system, using our scheduling algorithm, the resultant queue length process of all users over the time interval $t\in[0, T]$ is

\[
\textbf{Q}(t) = \left(Q_1(t), \cdots, Q_N(t)\right).
\]

\noindent
The other functions of the system over the same interval are given by 

\begin{eqnarray}
&\textbf{F}(t) = \left(F_1(t),\cdots,F_N(t)\right), \\
&\hat{\textbf{F}}(t) = \left(\hat{F}_1(t),\cdots,\hat{F}_N(t)\right), \\
&\textbf{G}(t) = \left(G_1(t), \cdots, G_M(t)\right), \\
&\hat{\textbf{G}} = \left(\hat{G}_m^i(t), m\in[1,\cdots, M], i\in[1,\cdots, N]\right).
\end{eqnarray}


The state of the system is uniquely determined by the initial state $\textbf{Q}(0)$. Functions $\textbf{F}$ and $\textbf{G}$ drive the system, and $\hat{\textbf{F}}$, $\hat{\textbf{G}}$ are determined by the scheduling algorithm and vice versa. The relationships of these functions over the time interval $\left[0,T\right]$ are given as follows.

\begin{eqnarray}
\label{eq:orig-rel1} & Q_i(T) = Q_i(0) + F_i(T) - {\hat{F}}_i(T), \forall{i}, \\
\label{eq:orig-rel2} & G_m(T) = \sum_i^N{\hat{G}}_m^i(T), \forall{m}.
\end{eqnarray}

Then, for any non-negative integer $B$, we define the scaled processes of the system are

\begin{eqnarray}
\label{sec:scaled-procs}
\nonumber &\left(\textbf{f}^B(t), \hat{\textbf{f}}^B(t), \textbf{g}^B(t), \hat{\textbf{g}}^B(t)\right) = \\
&\left(\frac{\textbf{F}(Bt)}{B}, \frac{\hat{\textbf{F}}(Bt)}{B}, \frac{\textbf{G}(Bt)}{B}, \frac{\hat{\textbf{G}}(Bt)}{B}\right).
\end{eqnarray}

For any given $T > 0$, let $\Psi_T$ denote the space of mappings from $\left[0,T\right]$ to $\left({\Re}^M, {\Re}^N\right)$. Since the arrival process of users follow poisson distribution, it is well known that the scaled arrival process of user $i$ at a particular time instant $t$, $f_i^B(t)$ has a sample path LDP with the following rate function

\begin{equation}
L_i(\dot{f_i}(t)) = \sup_{\theta \ge 0}\left[{\theta}\lambda_i - \mbox{log}Ee^{\theta\dot{f_i}(t)}\right],
\end{equation}

\noindent
where $L_i(.)$ is a convex lower semi-continuous function on $[0, \infty)$, and has the following properties

\[
L_i(\lambda_i) = 0, L_i(\xi) > 0~\mbox{for }\xi\neq\lambda_i, L_i(\xi)/\xi\to\infty \mbox{ as }\xi\to\infty. 
\]

Since the arrival processes of users are independent of each other, there exists a single decay rate function combining all users in the large deviation setting. This is because the decay rate of independent processes are additive. Hence, the combined rate function of all users at a particular time instant $t$ can be written as

\begin{equation}
\label{eq:prop1}
L_{(\textbf{f})}(\dot{\textbf{f}}(t)) = \sum_{i=1}^NL_i(\dot{f_i}(t)).
\end{equation}  

On the other hand, since the scaled channel state process $\textbf{g}^B(.)$ follows i.i.d. in the time domain, the rate function of this process can be given by

\begin{equation}
\label{eq:prop2}
L_{(\textbf{g})}(\dot{\textbf{g}}(t)) = H(\dot{\textbf{g}}(t)||\textbf{p}),
\end{equation}

\noindent
where $H(\bm{\gamma}||\textbf{p}) = \sum_{m=1}^M{\gamma}_m\mbox{log}\frac{{\gamma}_m}{p_m}$, and $\bm{\gamma} = [{\gamma}_1, \cdots, {\gamma}_M]\in\textit{P}_M$. Here, $\textit{P}_M\in{\Re}^M$ such that $\sum_{m=1}^M{\gamma}_m=1$ is satisfied. Now, the sequence of scaled processes $(\textbf{f}^B(.), \textbf{g}^B(.))$ are known to satisfy the sample path LDP over the interval $[0, T]$ with the rate function

\[
J_T(\textbf{f}^B(.), \textbf{g}^B(.)) = \begin{cases}\displaystyle\int_0^T\left[L_{(\textbf{f})}(\dot{\textbf{f}^B}(t)) + L_{(\textbf{g})}(\dot{\textbf{g}^B}(t))\right]dt,~\mbox{if }(\textbf{f}^B(.) \textbf{g}^B(.))\in AC, \\
\infty~\mbox{otherwise}\end{cases},
\]

\noindent
where $AC$ denotes the set of absolute continuous functions in $\Psi_T$. The intuitive interpretation of this function is that for any set $\lceil$ of trajectories in $\Psi_T$, the following relation holds:

\begin{eqnarray}
\nonumber-\inf_{(\textbf{f}^B(.), \textbf{g}^B(.))\in{\lceil}^o}J_T(\textbf{f}^B(.), \textbf{g}^B(.)) \ge \displaystyle\liminf_{B{\to}\infty}\frac{1}{B}\mbox{log}\textbf{P}\left[(\textbf{f}^B(.), \textbf{g}^B(.))\in\lceil\right] \\
\nonumber\ge \displaystyle\limsup_{B{\to}\infty}\frac{1}{B}\mbox{log}\textbf{P}\left[(\textbf{f}^B(.), \textbf{g}^B(.))\in\lceil\right] \\
\nonumber\ge -\inf_{(\textbf{f}^B(.), \textbf{g}^B(.))\in\bar{\lceil}}J_T(\textbf{f}^B(.), \textbf{g}^B(.)),
\end{eqnarray}

\noindent
here ${\lceil}^o$ and $\bar{\lceil}$ are the interior and closure of set $\lceil$, respectively. For $B=1,2,\cdots,\infty$, we can have different sequences of $\textbf{f}^B(.)$, $\textbf{g}^B(.)$, and $\textbf{q}^B(.)$. For any value of $B$, it is easy to verify that $\textbf{f}^B(.)$, $\textbf{g}^B(.)$, and $\textbf{q}^B(.)$ are Lipschitz-continuous, and their derivatives exist. As $B\to\infty$, there must exist a sequence of $\textbf{f}^B(.)$, $\textbf{g}^B(.)$, and $\textbf{q}^B(.)$ that converge uniformly over the interval $[0,T]$. Since the arrival scaled process $\textbf{f}^B(.)$ and the channel rate process $\textbf{g}^B(.)$ follow some known distribution, they follow some known sample path LDP. The goal of this work is to use the known sample path LDP of $\left(\textbf{f}^B(.), \textbf{g}^B(.)\right)$ to characterize that of $\textbf{q}^B(.)$ and the decay rate of the queue overflow probability.

\section{The Bounds on the Decay Rate of the Probability of $\left[\max_{1{\le}i{\le}N}Q_i(T) \ge B\right]$}
\label{sec:up-low-bound}

In this section, followed by the justification, we derive an upper bound $I_{opt}$ on $I^{UP}$ in (\ref{eq:eqn3}). Then, we provide a general lower bound of this on $I^{LOW}$. 

\subsection{The Upper Bound}

No matter the scheduling algorithm, $I_{opt}$ is the decay rate for the probability that the stationary backlog process $\textbf{Q}(.)$ exceeds a certain threshold after a certain time. For any arrival rate vector $\textbf{y}\in{\Re}^N$ and channel state probability vector $\bm{\gamma}\in\textit{P}_M$, define the following optimization problem

\begin{eqnarray}
&w(\textbf{y}, \bm{\gamma}) = \displaystyle\inf_{[{\phi}_m^i]} \displaystyle\max_{1 \le i \le N}\left[y_i - \sum_{m=1}^M{\gamma}_m{\phi}_m^iF_m^i\right]^+ \\
\nonumber &s.t. \sum_{i=1}^N{\phi}_m^i = 1,~\forall{m} \\
\nonumber & {\phi}_m^i \ge 0, \forall{i},\forall{m},
\end{eqnarray}

\noindent
where ${\phi}_m^i$ is the long term fraction of time that user $i$ is served when the channel state is $m$. Since $\textbf{f}(t)=\textbf{y}t$ and $\textbf{g}(t) = \bm{\gamma}t$, $\left[y_i - \sum_{m=1}^M\gamma_m{\phi}_m^iF_m^i\right]^+$ indicates the long term growth rate of the backlog of user $i$. If $\textbf{q}(0)=0$, $w(\textbf{y}, \bm{\gamma})$ is the minimum rate of growth of the backlog of the largest queue. Now, define

\[
I_{opt} = \inf_{\textbf{y}\in{\Re}^N, \bm{\gamma}\in{\Re}^M|w(\textbf{y}, \bm{\gamma}) \ge 0}\frac{L_{(\textbf{f})}(\textbf{y})+L_{(\textbf{g})}(\bm{\gamma})}{w(\textbf{y}, \bm{\gamma})}.
\]

\noindent \textbf{Proposition 1:} The stationary backlog process $\textbf{Q}(.)$ achieved by any scheduling algorithm satisfies the following relation

\[
\liminf_{B \to \infty}\frac{1}{B}\mbox{Pr}\left(\max_{1 \le i \le N}Q_i(0) \ge B \right) \ge -I_{opt}.
\]

\textit{Proof:} From the definition of $w(\textbf{y}, \bm{\pi})$, it implies that $w(.)$ provides a lower bound on the backlog of the largest queue, i.e.,

\begin{equation}
\label{eq:eqn3}
\max_{1 \le i \le N}q_i^B(T) \ge Tw(\textbf{f}^B(T)/T, \textbf{g}^B(T)/T),~\forall{(\textbf{f}^B(.), \textbf{g}^B(.)) \in {\Psi}_T}.
\end{equation}


For any $\delta > 0$, we can find $\left({\textbf{y}}_{\delta} \in {\Re}^N, {\bm{\gamma}}_{\delta} \in \textit{P}_M| w\left({\textbf{y}}_{\delta}, {\bm{\gamma}}_{\delta}\right) > 0\right)$ such that $(L_{(\textbf{g})}({\bm{\gamma}}_{\delta}) + L_{(\textbf{f})}({\textbf{y}}_{\delta}))/w({\textbf{y}}_{\delta}, {\bm{\gamma}}_{\delta}) < I_{opt}$. For any $t \ge 0$, we know that ${\textbf{f}}_{\delta}(t)={\textbf{y}}_{\delta}t$ and ${\textbf{g}}_{\delta}(t)={\bm{\gamma}}_{\delta}t$. 

Find $(\textbf{f}(.),\textbf{g}(.))\in{\Psi}_T({\textbf{f}}_{\delta}(.), {\textbf{g}}_{\delta}(.), \epsilon)\subseteq{\Psi}_T$ such that $\sup_{t{\in}[0,T]}||(\textbf{f}(t), \textbf{g}(t))-(\textbf{f}_{\delta}(t), \textbf{g}_{\delta}(t))|| < \epsilon$. It implies that $||(\textbf{f}^B(T)/T, \textbf{g}^B(T)/T)-(\textbf{y}_{\delta},\bm{\gamma}_{\delta})|| < {\epsilon}/T$ for $(\textbf{f}^B(.), \textbf{g}^B(.)){\in}{\Psi}_T({\textbf{f}}_{\delta}(.), {\textbf{g}}_{\delta}(.), \epsilon), B > 0$. Since $\max_{1{\le}i{\le}N}\left[y_i-\sum_{m=1}^M{\gamma}_m{\phi}_m^iF_m^i\right]^+$ is a continuous function with respect to $(\textbf{y}, \bm{\gamma})$ for any $\{{\phi}_m^i,\forall{m}, \forall{i}\}$, we can write

\begin{equation}
Tw\left(\frac{\textbf{f}^B(T)}{T}, \frac{\textbf{f}^B(T)}{T}\right) \ge Tw\left(\textbf{y}_{\delta}, \bm{\gamma}_{\delta}\right) - \epsilon{\epsilon}_1,
\end{equation}

\noindent
where ${\epsilon}_1$ is a small number and $> 0$. Now, if we define $T = (1 + \epsilon{\epsilon}_1)/w({\textbf{y}}_{\delta}, {\bm{\gamma}}_{\delta})$, from (\ref{eq:eqn3}), this in turn implies that

\begin{equation}
\displaystyle\max_{1 \le i \le N}q_i^B(T) \ge Tw({\textbf{y}}_{\delta}, {\bm{\gamma}}_{\delta}) -\epsilon{\epsilon}_1 = 1.
\end{equation}

\noindent
Therefore

\begin{eqnarray}
\nonumber &\mbox{Pr}\left(\max_{1 \le i \le N}Q_i(0) \ge B\right) = \mbox{Pr}\left(\max_{1 \le i \le N}Q_i(BT) \ge B\right) \\
\nonumber & = \mbox{Pr}\left(\max_{1 \le i \le N}q_i^B(T) \ge 1\right) \\
\nonumber & \approx \mbox{Pr}\left((\textbf{f}^B(.), \textbf{g}^B(.)) \in {\Psi}_T((\textbf{f}^B_{\delta}(.), \textbf{g}^B_{\delta}(.)), \epsilon)\right). 
\end{eqnarray}

\noindent
By the LDP definition of $(\textbf{f}^B(.), \textbf{g}^B(.))$, we have

\begin{eqnarray}
\nonumber &\displaystyle\liminf_{B \to \infty}\frac{1}{B}\mbox{log}\mbox{Pr}\left(\max_{1 \le i \le N}Q_i(T) \ge B\right) \\
\nonumber & \approx \displaystyle\liminf_{B \to \infty}\frac{1}{B}\mbox{log}\mbox{Pr}\left[(\textbf{f}^B(.), \textbf{g}^B(.)) \in {\Psi}_T((\textbf{f}^B_{\delta}(.), \textbf{g}^B_{\delta}(.)), \epsilon)\right] \\
\nonumber & \ge -\displaystyle\inf_{(\textbf{f}(.), \textbf{g}(.)) \in {\Psi}_T({\textbf{f}}_{\delta}, {\textbf{g}}_{\delta}, \epsilon)}{\int}_0^T\left[L_{(\textbf{f})}(\dot{\textbf{f}}(t)) + L_{(\textbf{g})}(\dot{\textbf{g}}(t))\right]dt \\
\nonumber & \ge -{\displaystyle\int}_0^T\left[L_{(\textbf{f})}(\dot{\textbf{f}}_{\delta}(t)) + L_{(\textbf{g})}(\dot{\textbf{g}}_{\delta}(t))\right]dt \\
\nonumber & = -T\left[L_{(\textbf{f})}({\textbf{y}}_{\delta}) + L_{(\textbf{g})}({\bm{\gamma}}_{\delta})\right] \\
\nonumber & = -\left(1 + \epsilon{\epsilon}_1\right)(I_{opt} + \delta).
\end{eqnarray}

Since $\delta$, $\epsilon$ and ${\epsilon}_1$ are arbitrarily small, we can conclude the proof of \textit{Proposition 1}.

\subsection{The General Lower Bound}

The large deviation philosophy implies that rare events occur in the most likely way. Hence, the probability of the queue overflow is determined by the smallest cost among all sample paths that overflow. So, the decay rate of the queue overflow probability is the minimum cost among all fluid sample paths that overflow. To prove the lower bound of this decay rate, we consider that the system starts at time 0, instead of considering the entities at the stationary system. Consider a certain time $T > 0$. Let ${\lceil}_T$ denote the set of fluid sample paths $\left(\textbf{f}(.), \textbf{g}(.), \textbf{q}(.)\right)$ on the interval $[0,T]$ such that $\textbf{q}(0) = 0$ and $\max_{1 \le i \le N}q_i(T) \ge 1$. \textit{Proposition 2} states the lower bound of the decay rate that the largest queue overflows.

\noindent
\textbf{Proposition 2:}  Given the system described above, we have the following lower bound

\begin{eqnarray}
\label{eq:eqn4}
& \nonumber \displaystyle\limsup_{B \to \infty}\frac{1}{B}\mbox{log}\mbox{Pr}\left[\textbf{q}^B(T) \ge 1\right] \\
& \le - \displaystyle\inf_{(\textbf{f}(.), \textbf{g}(.), \textbf{q}(.)) \in {\lceil}_T}\displaystyle\int_0^T[L_{(\textbf{f})}(\dot{\textbf{f}}(t)) + L_{(\textbf{g})}(\dot{\textbf{g}}(t))]dt + {\delta}T.
\end{eqnarray}

\textit{Proof: } For user $i$, let denote the maximum possible value of arrival rate is ${\Lambda}_1$. We choose a large integer $K_1$, and divide the interval $[0, {\Lambda}_1]$ into $K_1$ (integer number) subintervals such that the length of each subinterval is $\varsigma = {\Lambda}_1/K_1$. Hence, $k{\in}[1,\cdots,K_1]$th subinterval can be named as $[(k-1)\varsigma, k\varsigma)$. We call each granular subinterval as ``bin''. The value of $K_1$ is such that the following condition is satisfied for all bins of user $i$.

\begin{equation}
\label{eq:eqn5}
\mbox{max}\{|L_i(y_1) - L_i(y_2)|: L_i(y_1), L_i(y_2) < \infty, y_1, y_2 \in [(k-1)\varsigma, k\varsigma)\} < {\epsilon}/N.
\end{equation}

We ensure the property in (\ref{eq:eqn5}) uniformly over all bins for all users. If necessary, we increase/decrease the value of $K_i, i{\in}[1,\cdots,N]$. Then, we divide the simplex of all vectors representing the probability distributions $\bm{\gamma}$ on the set of channel states into $K_{N+1}$ (integer number) non-intersecting subsets (``bins'') such that the oscillation (difference between the maximum and minimum) of $L_{(\textbf{g})}(\bm{\gamma})$ within the closure of each bin is at most $\epsilon$. If necessary, we increase/decrease the value of $K_{N+1}$ so that this condition is satisfied uniformly over all bins.

For any $B > 0$, and for the time interval $[0, T]$, let a vector function $\textbf{h}^B(.) = (\textbf{f}_i^B(.), i= 1,\cdots,N; \textbf{g}^B(.))$. Each component of $\textbf{h}^B(.)$ has a constant non-negative derivative in each of the time-subintervals of $[0,T]$. We know that the derivative of $f_i(.)$ ($\dot{\textbf{f}}_i(.)$) resides in each of the $K_i$ bins (described above), and the derivative of $\textbf{g}(.)$ ($\dot{\textbf{g}}(.)$) is in each of the $K_{N+1}$ bins. Consequently, the derivatives of the function $\textbf{h}^B(.)$ over the time interval $[0,T]$ has $[\prod_{i=1}^{N+1}K_i]^T$ possible combinations. For any $B > 0$, consider one fixed aggregate bin $B_{ab}$, of which $[\xi_1, \xi_2]$ is the bin for the component $f_i^B(t)$ of $\textbf{h}^B(t)$. Hence, the LDP definition of $\textbf{f}_i^B(.)$ at time instant $t$ implies that

\begin{equation}
\label{eq:eqn6}
\mbox{log}\mbox{Pr}\left[f_i^B(t) - f_i^B(t-1)\right] \le -\displaystyle\min_{\xi\in[\xi_1,\xi_2]}L_i(\xi) + \frac{\epsilon}{N}.
\end{equation}

The property in~(\ref{eq:eqn6}) is applicable for other component of $\textbf{h}^B(.)$ as well. Since the components of $\textbf{h}^B(.)$ are independent processes, the decay rate of the derivative of $\textbf{h}^B(.)$ is the sum value of the decay rate of the derivatives of its components. Therefore, combining all components and according to the definition of LDP, we can write

\begin{eqnarray}
\label{eq:eqn7}
\nonumber & \mbox{log}\mbox{Pr}\left[\textbf{h}^B(.) \in B_{ab}\right] \le \\
& - \displaystyle\int_0^T\left[L_{(\textbf{f})}(\dot{\textbf{f}}(t)) + L_{(\textbf{g})}(\dot{\textbf{g}}(t))\right] + 2{\epsilon}T.
\end{eqnarray}

For $\forall{B\in[1,\cdots,\infty]}$, we set $\textbf{q}^B(0) = 0$. Then, for $T > 0$ and for any $B$, ${\lceil}^B$ is the set of arrival and channel rate processes $(\textbf{f}^B(.), \textbf{g}^B(.))$ such that the corresponding backlog process satisfies $\max_{1 \le i \le N}q_i^B(T) \ge 1$. Obviously, each component of ${\lceil}^B$ is the instance of vector function $\textbf{h}^B(.)$. Consequently, we have

\begin{eqnarray}
\nonumber \displaystyle\limsup_{B \to \infty}\frac{1}{B}\mbox{log}\mbox{Pr}\left[(\textbf{f}^B(.), \textbf{g}^B(.)) \in {\lceil}^B\right] \\
\le \displaystyle\limsup_{B \to \infty}\frac{1}{B}\mbox{log}\mbox{Pr}\left[(\textbf{f}^B(.), \textbf{g}^B(.)) \in \cup_{B=1}^{\infty}{\lceil}^B\right]
\end{eqnarray}  

\noindent
By the LDP definition for $\textbf{h}^B(.)$ ($(\textbf{f}^B(.), \textbf{g}^B(.))$) in~(\ref{eq:eqn7}), we have

\begin{eqnarray}
\nonumber \displaystyle\limsup_{B \to \infty}\frac{1}{B}\mbox{log}\mbox{Pr}\left[(\textbf{f}^B(.), \textbf{g}^B(.)) \in \cup_{B=1}^{\infty}{\lceil}^B\right] \\
\le - \inf_{(\textbf{f}(.), \textbf{g}(.)) \in \overline{\cup_{B=1}^{\infty}{\lceil}^B}}\displaystyle\int_0^T\left[L_{(\textbf{f})}(\dot{\textbf{f}}(t)) + L_{(\textbf{g})}(\dot{\textbf{g}}(t))\right]dt + 2{\epsilon}_0T.
\end{eqnarray}

\noindent
Since ${\lceil}^B{\subseteq}\cup_{B=1}^{\infty}{\lceil}^B$, it turns out

\begin{eqnarray}
\label{eq:eqn8}
\nonumber \displaystyle\limsup_{B \to \infty}\frac{1}{B}\mbox{log}\mbox{Pr}\left[(\textbf{f}^B(.), \textbf{g}^B(.)) \in {\lceil}^B\right] \\
\le - \inf_{(\textbf{f}(.), \textbf{g}(.)) \in \overline{\cup_{B=1}^{\infty}{\lceil}^B}}\displaystyle\int_0^T\left[L_{(\textbf{f})}(\dot{\textbf{f}}(t)) + L_{(\textbf{g})}(\dot{\textbf{g}}(t))\right]dt + 2{\epsilon}_0T.
\end{eqnarray}

To prove \textit{Proposition 2}, it is sufficient to show that the right hand side of (\ref{eq:eqn6}) is no greater than that of (\ref{eq:eqn8}). For each $n \ge 1$, we can find $(\textbf{u}_n(.), \textbf{v}_n(.))\in\overline{\cup_{B=1}^{\infty}{\lceil}^B}$ such that

\begin{eqnarray}
\label{eq:eqn9}
&\nonumber \displaystyle\int_0^T\left[L_{(\textbf{f})}(\dot{\textbf{u}}_n(t)) + L_{(\textbf{g})}(\dot{\textbf{v}}_n(t))\right]dt - 2{\epsilon}_nT \\ 
&< \displaystyle\inf_{(\textbf{f}(.), \textbf{g}(.)) \in \overline{\cup_{B=1}^{\infty}{\lceil}^B}}\displaystyle\int_0^T\left[L_{(\textbf{f})}(\dot{\textbf{f}}(t)) + L_{(\textbf{g})}(\dot{\textbf{g}}(t))\right]dt - 2{\epsilon}_0T + \frac{1}{n}.
\end{eqnarray}

Since $(\textbf{u}_n, \textbf{v}_n)$ is equicontinuous, we can find a subsequence that has a uniform derivative on the time-interval $[0, T]$. Let $(\textbf{u}_n^*, \textbf{v}_n^*)$ be its limit, which implies that $\lim_{n\to\infty}(\textbf{u}_n(.), \textbf{v}_n(.)) = (\textbf{u}_n^*, \textbf{v}_n^*)$. Since the cost functions in (\ref{eq:prop1}) and (\ref{eq:prop2}) are lower semi-continuous, we can have

\begin{eqnarray}
\label{eq:eqn10}
&\nonumber \displaystyle\liminf_{n\to\infty}\displaystyle\int_0^T\left[L_{(\textbf{f})}(\dot{\textbf{u}}_n(t)) + L_{(\textbf{g})}(\dot{\textbf{v}}_n(t))\right]dt - 2{\epsilon}_nT \\
& \ge \displaystyle\liminf_{n\to\infty}\displaystyle\int_0^T\left[L_{(\textbf{f})}(\dot{\textbf{u}}_n^*(t)) + L_{(\textbf{g})}(\dot{\textbf{v}}_n^*(t))\right]dt - 2{\epsilon}_n^*T.
\end{eqnarray}

The process of obtaining $(\textbf{u}_n^*(.), \textbf{v}_n^*(.))$ from $(\textbf{u}_n(.), \textbf{v}_n(.))\in\overline{\cup_{B=1}^{\infty}{\lceil}^B}$ is described as follows. From $(\textbf{u}_n(.), \textbf{v}_n(.))$, we can find a series of functions $(\textbf{u}_{n,m}(.), \textbf{v}_{n,m}(.)), m=1,\cdots,\infty$ such that $\lim_{m\to\infty}(\textbf{u}_{n,m}(.), \textbf{v}_{n,m}(.)) = (\textbf{u}_n^*(.), \textbf{v}_n^*(.))$. From subscript $m$, we can choose $m+1$ such that $\sup_{t\in[0,T]}||(\textbf{u}_{n,m+1}(t), \textbf{v}_{n,m+1}(t)) - (\textbf{u}_{n,m+1}(t-1), \textbf{v}_{n,m+1}(t-1))|| < \sup_{t\in[0,T]}||(\textbf{u}_{n,m}(t), \textbf{v}_{n,m}(t)) - (\textbf{u}_{n,m}(t-1), \textbf{v}_{n,m}(t-1))||$ is satisfied. 

For each $n$, let $\textbf{q}_n^*(.)$ be the corresponding backlog process for $(\textbf{u}_n^*(.), \textbf{v}_n^*(.))$. Similar to $(\textbf{u}_n^*(.), \textbf{v}_n^*(.))$, $\textbf{q}_n^*(.)$ is equi-continuous and has a uniform non-negative derivative on the interval $[0,T]$. From the construction, $\textbf{q}_n^*(0) = 0$ and $\{\max_{1{\le}i{\le}N}q_i(T)\}_n^* = 1$ are true. Hence, $(\textbf{u}_n^*(.), \textbf{v}_n^*(.), \textbf{q}_n^*(.))$ is in ${\lceil}_T$. Consequently, for sufficiently small values of $\epsilon_n$, $\epsilon_n^*$ and $\delta$, we can write

\[
\displaystyle\int_0^T\left[L_{(\textbf{f})}(\dot{\textbf{u}}_n^*(t)) + L_{(\textbf{g})}(\dot{\textbf{v}}_n^*(t))\right]dt - 2{\epsilon}_n^*T \ge \displaystyle\inf_{(\textbf{f}(.), \textbf{g}(.), \textbf{q}(.)) \in {\lceil}_T}\displaystyle\int_0^T[L_{(\textbf{f})}(\dot{\textbf{f}}(t)) + L_{(\textbf{g})}(\dot{\textbf{g}}(t))]dt - {\delta}T.
\]

\noindent
Finally, from (\ref{eq:eqn9}) and (\ref{eq:eqn10}), we can conclude that

\begin{eqnarray}
& \nonumber \displaystyle\inf_{(\textbf{f}(.), \textbf{g}(.)) \in \overline{\cup_{B=1}^{\infty}{\lceil}^B}}\displaystyle\int_0^T\left[L_{(\textbf{f})}(\dot{\textbf{f}}(t)) + L_{(\textbf{g})}(\dot{\textbf{g}}(t))\right]dt - 2{\epsilon}_0T \\
& \nonumber \ge \displaystyle\liminf_{n\to\infty} \displaystyle\int_0^T\left[L_{(\textbf{f})}(\dot{\textbf{u}}_n(t)) + L_{(\textbf{g})}(\dot{\textbf{v}}_n(t))\right]dt - 2{\epsilon}_nT \\
& \ge \displaystyle\inf_{(\textbf{f}(.), \textbf{g}(.), \textbf{q}(.)) \in {\lceil}_T}\displaystyle\int_0^T[L_{(\textbf{f})}(\dot{\textbf{f}}(t)) + L_{(\textbf{g})}(\dot{\textbf{g}}(t))]dt - {\delta}T.
\end{eqnarray}

\noindent
Thus, \textit{Proposition 2} is proved.

\section{Sample Path Properties to the Largest Queue Overflow}
\label{sec:prop}

In this section, we will show the properties of the path to $\left[\max_{1{\le}i{\le}N}Q_i(T) \ge B\right]$ achieved by our scheduling algorithm. In this context, we would like to study the relationship of the system components and their derivatives in granular time-interval. To understand the behavior of the system, scaled processes are already derived and given by (\ref{sec:scaled-procs}). To interconnect the outcome of scheduling policy with these scaled processes, we study the scheduling rule at time instant $Bt$ for any $B > 0$ and $t > 0$. After re-arranging the scheduling rule for user $i$ at time instant $Bt$, we have $- e^{\left[\rho_1 + \rho_2 - \frac{F_m^i}{\max_{i{\in}\textbf{N}}F_m^i}\right]}e^{-\frac{Q_i(Bt)}{Q^{th}}}$ if the channel state is $m$ at time instant $Bt$. Once scaled over $B$, it appears $- e^{\left[\rho_1 + \rho_2 - \frac{F_m^i}{\max_{i}F_m^i}\right]}\frac{1}{B}e^{-B\frac{q_i^B(t)}{Q^{th}}} \approx - e^{\left[\rho_1 + \rho_2 - \frac{F_m^i}{\max_{i}F_m^i}\right]}e^{-\frac{q_i^B(t)}{Q^{th}}}$. Since the scaled queue length part of our scheduling rule is proportional to some real number multiple of the original non-scaled queue length, it is sufficient to find the relationship between the system components and the scheduling policy at the scaled time instant $t$ instead of further refinement. \textit{Lemma 1} summarizes these relationships followed by the corresponding proof. In a non-overloaded state of the system, we provide the structure of the sample path to overflow event over a finite time interval in \textit{Theorem 1}, and then in \textit{Theorem 2}, we prove that the value of $Q_{th}$ of our scheduling rule affects the decay rate of the largest queue overflow probability.

\noindent
\textbf{Lemma 1:} At the scaled time instant $t{\in}[0,T]$, the following derivatives exist and are finite

\begin{eqnarray}
\label{eq:derivs}
\nonumber & \bm{\lambda}(t) = \frac{d}{dt}\textbf{f}(t);~~~~\bm{\gamma}(t) = \frac{d}{dt}\textbf{g}(t), \\
\nonumber & \bm{\mu}(t) = \frac{d}{dt}\hat{\textbf{f}}(t);~~~~\frac{d}{dt}\hat{\textbf{g}}(t), \\
& \frac{d}{dt}\textbf{q}(t);~~~~ \frac{d}{dt}q_*(t),~\mbox{where }q_*(t) = \max_iq_i(t).
\end{eqnarray}

\noindent
Furthermore, the following relationships hold among the system components, their derivatives and our scheduling policy 

\begin{eqnarray}
\label{eq:rel1} &\frac{d}{dt}\textbf{q}(t) = \bm{\lambda}(t) - \bm{\mu}(t), \\
\label{eq:rel2} &\frac{d}{dt}q_*(t) = \frac{d}{dt}q_i(t), \mbox{ for user }i\mbox{ such that }q_i(t) = q_*(t), \\
\label{eq:rel3} &\mu_i(t) = \displaystyle\sum_mF_m^i\frac{d}{dt}{\hat{g}}_m^i(t), \forall{i},\\
\label{eq:rel4} &\gamma_m(t) = \displaystyle\sum_i\frac{d}{dt}{\hat{g}}_m^i(t), \forall{m}, \\
\label{eq:rel5} &\bm{\mu}(t) \in \displaystyle\argmax_{\textbf{v}\in\textbf{V}_{\bm{\gamma}(t)}}-e^{\left[\rho_1 + \rho_2 - \frac{v_i}{\max_iv_i}\right]}e^{-\frac{q_i(t)}{Q^{th}}}.
\end{eqnarray}

\textit{Proof: }For moderately large value of $B>0$, the sequence of scaled functions 
$(\textbf{f}^B(t), \hat{\textbf{f}}^B(t),$ $\textbf{g}^B(t), \hat{\textbf{g}}^B(t), \textbf{q}^B(t), q_*^B(t))$ converges to $(\textbf{f}(t), \hat{\textbf{f}}(t), \textbf{g}(t), \hat{\textbf{g}}(t), \textbf{q}(t), q_*(t))$ over the time interval $t\in[0,T]$. Since all functions are Lipschitz-continuous, their derivatives in (\ref{eq:derivs}) exist. The relations in (\ref{eq:rel1}), (\ref{eq:rel3}) and (\ref{eq:rel4}) are a simple consequence of the relations in (\ref{eq:orig-rel1}) and (\ref{eq:orig-rel2}). Since our objective is to study the path to the largest queue overflow event, we denote the largest queue at scaled time instant $t$ by $q_*(t)$. Since at scaled time instant $t$, $(d/dt)q_i(t), \forall{i}$ exists, the relation in (\ref{eq:rel2}) is satisfied consequently. Consider a scaled time-interval $[t-1, t]$. The unscaled version of this interval is $[B(t-1), Bt]$. During this unscaled time-interval, at any time instant, if the channel state $m$ is selected, user $i*$ is chosen for service based on the rule $i^*\in\argmax_{i}- e^{\left[\rho_1 + \rho_2 - \frac{F_m^i}{\max_{i}F_m^i}\right]}e^{-\frac{q_i^B(t)}{Q^{th}}}$. However, at its counterpart scaled $t$th time instant, the service rate of each user is the average service rate achieved over $B$ time slots. Given the channel state distribution over $B$ time slots $\bm{\gamma}(.)$, we can deduce a set which contains service rate of all users for all possible combinations, $\textbf{V}_{\bm{\gamma}(.)}$. Given this set, the scheduling policy at time instant $t$ converges to the relation in (\ref{eq:rel5}). The detailed proof is given in~\cite{Shakkottai00} for the EXP scheduling policy.

\noindent
\textbf{Lemma 2:} If $J_t(\textbf{f}, \textbf{g}) = L_{(\textbf{f})}(\dot{\textbf{f}}(t)) +  L_{(\textbf{g})}(\dot{\textbf{g}}(t))$, there exist some fixed constants $\epsilon > 0$ and $\delta > 0$ such that the following statement holds at any time $t{\in}[0,T]$ for a non-overloaded system

\[
\frac{d}{dt}J_t(\textbf{f}, \textbf{g}) \le \epsilon~\mbox{implies}~\frac{d}{dt}q_*(t) \le -\delta
\]

\textit{Proof:} Let $\bm{\phi}^*$ be the channel-user assignment matrix for the best possible non-overloaded system, and $\textbf{v}^*$ be the corresponding service rate vector such that $\textbf{v}^* = \{\sum_mp_m\phi_m^iF_m^i, \forall{i}, \phi_m^i\in\bm{\phi}^*\}$. At time instant $t$, consider a subset of users $\textbf{N}^*$, the users of which satisfy $q_i(t) = q_*(t)$. Let us assume that channel state $m$ is sampled at time instant $t$, such that $\gamma_m(t) > 0$ and $\phi_m^i = (\frac{d}{dt}\hat{g}_m^i)/\gamma_m(t)$.  Then, using an argument similar to that in~\cite{Shakkottai00} (at the end of Section $4.3$), we can establish the following fact

\[
\displaystyle\sum_{i{\in}\textbf{N}^*}\phi_{mi}(t) \ge \displaystyle\sum_{i{\in}\textbf{N}^*}\phi_{mi}^*.
\]

\noindent
Multiplying $F_m^i$ on both sides for user $i{\in}\textbf{N}^*$, we obtain

\begin{equation}
\label{eq:eqn11}
\displaystyle\sum_{i{\in}\textbf{N}^*}\phi_{mi}(t)F_m^i \ge \displaystyle\sum_{i{\in}\textbf{N}^*}\phi_{mi}^*F_m^i.
\end{equation}

Now, $\frac{d}{dt}J_t(\textbf{f}, \textbf{g}) \le \epsilon$ implies that $\bm{\lambda}(t)$ is close to $\bm{\lambda}$, and $\bm{\gamma}(t)$ is close to $\textbf{p}$. Furthermore assume that stochastic matrix $\bm{\phi}^*$ is the resultant outcome for the actual arrival rate distribution $\bm{\lambda}$ of the system. Therefore, for all sufficiently small values (i.e., $\epsilon > 0$) of $\frac{d}{dt}J_t(\textbf{f}, \textbf{g})$, we have

\[
\bm{\lambda}(t) < \bm{\lambda},~\mbox{and}~ \bm{\lambda} < \textbf{v}^*.
\]

\noindent
After multiplying $\gamma_m(t)$ for channel state $m$, and then summing the relation in~(\ref{eq:eqn11}) over all channel states, we obtain

\[
\displaystyle\sum_{i{\in}\textbf{N}^*}\mu_i(t) \ge \displaystyle\sum_{i{\in}\textbf{N}^*}v_i^*.
\]

\noindent
Sufficiently small values (i.e., $\epsilon > 0$) of $\frac{d}{dt}J_t(\textbf{f}, \textbf{g})$ further implies

\[
\displaystyle\sum_{i{\in}\textbf{N}^*}[\lambda_i(t) - \mu_i(t)] \le \sum_{i{\in}\textbf{N}^*}[\lambda_i - v_i^*]  < 0.
\]

At time instant $t{\in}[0,T]$, we know that $q_*(t)=\lambda_i(t) - \mu_i(t), \forall{i{\in}\textbf{N}^*}$. Hence, the statement of \textit{Lemma 2} is proved.

\noindent
\textbf{Lemma 3:} There exist fixed constants $\epsilon > 0$ and $\delta > 0$ such that the following statement holds for the time-interval $[0, T]$ at a non-overloaded state of the system

\[
J_T(\textbf{f}, \textbf{g}) - J_0(\textbf{f}, \textbf{g}) \le {\epsilon}T~\mbox{ implies that}~q_*(T) - q_*(0) \le -{\delta}T.
\]

\textit{Proof: }The proof is as same as \textit{Lemma 9.3} of~\cite{Stolyar08}.

\noindent
\textbf{Theorem 1:} In a non-overloaded state of the system, for any $\delta_1 > \delta_2 >  0$, and $T > 0$, let us denote

\[
K(\delta_1, \delta_2, T) = \mbox{inf}\{L_{(\textbf{f})}(\dot{\textbf{f}}(t))+L_{(\textbf{g})}(\dot{\textbf{g}}(t))| q_*(0)\le \delta_1~\mbox{and}~q_*(t) \le \delta_2,~\mbox{for}~t\in[0,T]\},
\]

\noindent
The aforementioned definition implies that

\[
\mbox{log}\displaystyle\sup_{q_*(0) \le \delta_1}\mbox{Pr}\left(\displaystyle\sup_{t\in[0,T]}q_*(T) \ge \delta_2\right) \le -K(\delta_1, \delta_2, T).
\]

\noindent

Based on the above definitions, given some fixed constant $\delta_1 > 0$, the value of $K(\delta_1, \delta_2, T)$ grows linearly with $T$. More precisely, for any $\delta_1 > 0$, there exists $\Delta > 0$ such that for a moderately large value of $T > 0$ and $\delta_2\in(0, \delta_1)$, $K(\delta_1, \delta_2, T) \ge {\Delta}T$.

\textit{Proof: }The proof of this theorem can be proved by the statement of \textit{Lemma 3}.

\noindent
\textbf{Theorem 2: }At any time instant, suppose $\bm{\lambda}$ and $\bm{\gamma}$ are the arrival rate and channel state vectors\footnote{For the sake of simplicity, we omit time instant $t$.}. Consequently, $\textbf{V}_{\bm{\gamma}}$ be the set of all possible service rate vectors for all users. Furthermore, we assume that $\bm{\lambda}{\notin}\textbf{V}_{\bm{\gamma}}$. If $\omega$ is the growth rate of the largest queue, and $\bm{\mu}$ is the service rate vector, the value of $\omega$ resides in the solution of the auxiliary optimization problem in~(\ref{eq:opt1}).

\begin{eqnarray}
\label{eq:opt1}
\nonumber &\displaystyle\max_{\textbf{q}}\displaystyle\min_{\textbf{v}\in\textbf{V}_{\bm{\gamma}}}\sum_i-e^{-q_i/Q^{th}}\left[\lambda_i - e^{\{-\frac{v_i}{\max_iv_i} + \rho_1 + \rho_2\}}\right] \\
&\mbox{subject to } \sum_{i}e^{-q_i/Q^{th}} \le A,~\mbox{is a fixed constant and }> 0. 
\end{eqnarray}

\noindent
The optimization problem in (\ref{eq:opt1}) has the following properties

\noindent
(i) It has the following structure, which is equivalent to the rule of our scheduling algorithm.  

\begin{eqnarray}
\nonumber &\textbf{b}{\in}\displaystyle\argmax_{e^{-\frac{\textbf{v}}{\max_iv_i}}{\in}{\Re}_N}-e^{-q_i/Q^{th}}e^{\{-\frac{v_i}{\max_iv_i} + \rho_1 + \rho_2\}} \\
\nonumber &\mbox{such that }\lambda_i - b_i = \pi,~\pi~\mbox{is a positive number}, \\
&\mbox{where }\textbf{b}~\mbox{is the real number multiple of}~\textbf{v},~\bm{\mu} = \textbf{v},~\mbox{and}~\omega = \displaystyle\max_{1 \le i \le N}\lambda_i - \mu_i.
\end{eqnarray}

\noindent
(ii) The resultant value of $\omega$ is affected by the value of $Q_{th}$, and the optimal decay rate achieved by our scheduling algorithm can be obtained by tweaking this variable. 

\noindent
(iii) It is equivalent to the optimization problem in (\ref{eq:opt2}). And, the value of the problem in (\ref{eq:opt1}) is equal to $A$ multiple of the value in (\ref{eq:opt2}).

\begin{eqnarray}
\label{eq:opt2}
\nonumber &\displaystyle\min_{\textbf{v}{\in}\textbf{V}_{\bm{\gamma}}}\displaystyle\max_{1{\le}i{\le}N}\left[\lambda_i - e^{\{-\frac{v_i}{\max_iv_i} + \rho_1 + \rho_2\}}\right] \\
& \mbox{subject to }v_i \ge 0, \forall{i}.
\end{eqnarray}

\textit{Proof: }If we redefine the variables $-e^{-q_i/Q^{th}} = z_i, \forall{i}$, the optimization problem in (\ref{eq:opt1}) can be written as

\begin{eqnarray}
\label{eq:opt3}
\nonumber &\displaystyle\max_{\textbf{z}{\in}{\Re}^N}\displaystyle\min_{\textbf{v}{\in}\textbf{V}_{\bm{\gamma}}}\sum_iz_i\left[\lambda_i - e^{\{-\frac{v_i}{\max_iv_i} + \rho_1 + \rho_2\}}\right] \\
\nonumber &\mbox{subject to }\sum_iz_i \le A \\
&z_i \ge 0, \forall{i}.
\end{eqnarray}

Note the following property of the problem in (\ref{eq:opt3})

\begin{equation}
\label{eq:func1}
X(\textbf{z}) = \max_{\textbf{v}{\in}\textbf{V}_{\bm{\gamma}}}\sum_iz_ie^{\{-\frac{v_i}{\max_iv_i} + \rho_1 + \rho_2\}}.
\end{equation}

The aforementioned function in (\ref{eq:func1}) is convex. The inner problem (the min part) in (\ref{eq:opt3}) is then the concave function $\hat{X}(\textbf{z}) = \textbf{z}\bm{\lambda} - X(\textbf{z})$, and then we can write the convex problem in (\ref{eq:opt4}) as

\begin{equation}
\label{eq:opt4}
\displaystyle\max_{\textbf{z}{\in}{\Re}^N}\hat{X}(\textbf{z}),
\end{equation} 

\noindent
subject to the constraints in (\ref{eq:opt3}). Taking the Lagrangian of the problem in (\ref{eq:opt4}) over \textbf{z}, we have

\begin{equation}
L(\textbf{z}) = \hat{X}(\textbf{z}) - \alpha(\sum_iz_i - A) + \sum_i\beta_iz_i,
\end{equation}

\noindent
where $\pi$ and $\beta_i, \forall{i}$, are the Lagrange multipliers. For any optimal solution $\textbf{z}^*$ of the problem, there exist some fixed $\alpha$ and $\beta_i, \forall{i}$ for which the following conditions are satisfied

\begin{equation}
\bm{\lambda} - \textbf{b} + \left(-\pi + \beta_1,\cdots,-\pi + \beta_N\right) = 0,
\end{equation}

\noindent
where

\begin{equation}
\textbf{b} \in \argmax_{e^{-\frac{\textbf{v}}{\max_iv_i}}{\in}{\Re}_N}\textbf{z}e^{\{-\frac{\textbf{v}}{\max_iv_i} + \rho_1 + \rho_2\}}.
\end{equation}

This is because a vector $\textbf{b}$ is a subgradient of $X(\textbf{z})$ at point $\textbf{z}$ if and only if $\textbf{b} \in \displaystyle\argmax_{e^{-\frac{\textbf{v}}{\max_iv_i}}{\in}{\Re}_N}\textbf{z}e^{\{-\frac{\textbf{v}}{\max_iv_i} + \rho_1 + \rho_2\}}$. Due to the duality property, $\pi$ must be positive that implies $\lambda_i - b_i, \forall{i}$ are positive, and $\beta_i = 0, \forall{i}$ since $\textbf{z}^*$ is positive. Since $b_i = e^{-\frac{v_i}{\max_iv_i}},\forall{i}$, we can say that the resultant $v_i, \forall{i}$ is the real number multiple of $b_i, \forall{i}$. Consequently, from the definition, $b_i,\forall{i}$ can be renamed as the service rate vector $\bm{\mu}$. Furthermore, from the definition of the growth rate of the largest queue, $\omega = \max_{1 \le i \le N}\lambda_i - \mu_i$. This proves the statement of (i) in \textit{Theorem 2}. Consequently, at the unique optimal solution $\textbf{z}^*$, the value of the problem in (\ref{eq:opt3}) is

\[
\displaystyle\sum_iz_i^*\left[\lambda_i - e^{\{-\frac{\mu_i^*}{\max_i\mu_i^*} + \rho_1 + \rho_2\}}\right] = \displaystyle\sum_iz_i^*\left[\lambda_i - b_i\right] = A\pi. 
\]

Now, we want to see how the value of $Q_{th}$ affects the value of $\bm{\mu}$ that eventually affects the value of $\omega$. Intuitively, the larger the value of $Q_{th}$, the resultant $\textbf{b}$ is such that it gives more priority to the value of $e^{\{-\frac{v_i}{\max_iv_i} + \rho_1 + \rho_2\}}, \forall{i}$ instead of $\textbf{z}$. On the other hand, the smaller the value of $Q_{th}$, the resultant $\textbf{b}$ is such that it gives more priority to the value of \textbf{z}. Consequently, the values of $\bm{\mu}$ and $\omega$ are affected as well. To summarize, consider $Q_{th}^1$ and $Q_{th}^2$ are two possible values of $Q_{th}$, where $Q_{th}^1 \neq Q_{th}^2$. For $Q_{th} = Q_{th}^1$ and $Q_{th} = Q_{th}^2$, if the resultant values of $\omega$ are $\omega_1$ and $\omega_2$, respectively, we can say that $\omega_1 \neq \omega_2$. The optimal value of $\omega$ (the minimum one) resides in choosing the proper value of $Q_{th}$. Generally, the smaller the value of $Q_{th}$ (while $Q_{th} \neq 0$), the better the value of $\omega$. This concludes the statement of (ii) in \textit{Theorem 2}.

The proof of the statement in (iii) is as follows. We know that $[\lambda_i - e^{\{-\frac{v_i}{\max_iv_i} + \rho_1 + \rho_2\}}] \le \max_i[\lambda_i - e^{\{-\frac{v_i}{\max_iv_i} + \rho_1 + \rho_2\}}], \forall{i}$. Hence, the objective function part of (\ref{eq:opt3}) can be re-written as

\[
z_i\left[\lambda_i - e^{\{-\frac{v_i}{\max_iv_i} + \rho_1 + \rho_2\}}\right] \le z_i\displaystyle\max_{1 \le i \le N}\left[\lambda_i - e^{\{-\frac{v_i}{\max_iv_i} + \rho_1 + \rho_2\}}\right], \forall{i}.
\]

\noindent
Consequently, the maximum value of the problem in (\ref{eq:opt3}) (while satisfying its constraint $\sum_{i}z_i \le A$) appears to

\begin{eqnarray}
\label{eq:opt4}
\nonumber &A\displaystyle\min_{\textbf{v}{\in}\textbf{V}_{\bm{\gamma}}}\displaystyle\max_{1{\le}i{\le}N}\left[\lambda_i - e^{\{-\frac{v_i}{\max_iv_i} + \rho_1 + \rho_2\}}\right] \\
& \mbox{subject to }v_i \ge 0, \forall{i}.
\end{eqnarray}

\noindent
This concludes the statement of (iii) in \textit{Theorem 2}.

\section{Numerical Results}
\label{sec:sim-res}

\begin{table}
  \label{tab:F-param}
  \caption{Channel Capacity in Different States}
  \begin{center}
   \begin{tabular}{ | c || c | c | c | }
       \hline
       $F_m^i$ & $m=1$ & $m=2$ & $m=3$ \\ \hline \hline
       $i=1$ & 0 & 3 & 5 \\ \hline
       $i=2$ & 0 & 9 & 0 \\ \hline
       $i=3$ & 0 & 9 & 1 \\ \hline
       $i=4$ & 0 & 9 & 1 \\ 
       \hline
    \end{tabular}
  \end{center}
\end{table}

\begin{table}
  \label{tab:I-opt-phi}
  \caption{$\phi_m^i$ of the Optimal Decay Rate ($I_{opt}$ = 0.4518)}
  \begin{center}
   \begin{tabular}{ | c || c | c | c | }
       \hline
       & $m=1$ & $m=2$ & $m=3$ \\ \hline \hline
       $i=1$ & 0.25 & 0.3137 & 1 \\ \hline
       $i=2$ & 0.25 & 0.2288 & 0 \\ \hline
       $i=3$ & 0.25 & 0.2288 & 0 \\ \hline
       $i=4$ & 0.25 & 0.2288 & 0 \\ 
       \hline
    \end{tabular}
  \end{center}
\end{table}

In this section, we will provide simulation results to verify the analytical results given in the earlier sections. We consider that the simulated system has $4$ users (i.e., $N=4$) and $3$ channel states (i.e., $M=3$). Data packet arrival rate for all users are considered as $1$ (i.e., $\lambda_i=1, \forall{i}$). The probabilities of the channel states are $p_1=0.3, p_2=0.6$ and $p_3=0.1$. The capacity for each user on each channel state is given in Table I. 

\begin{figure}
   \begin{center}
    \includegraphics[width=0.6\columnwidth]{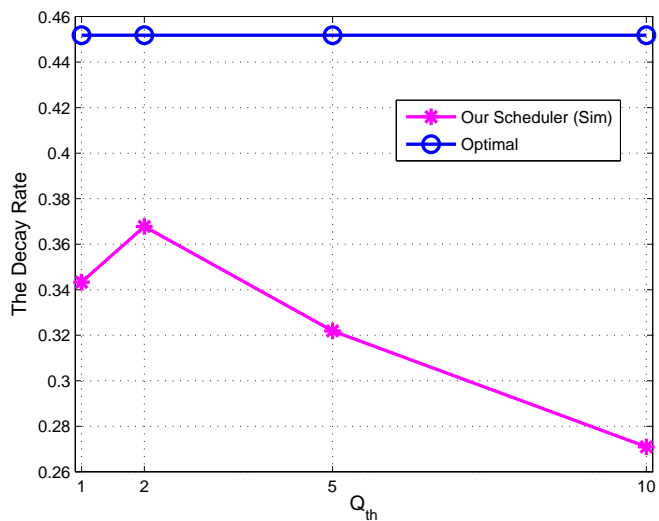}
    \caption{Decay rate comparison between our scheduler ($B = 40$) and the optimal one.}
    \label{fig:our-DR}
  \end{center}
\end{figure}

\begin{table}
\centering
\subtable[$Q_{th} = 2$]
{\begin{tabular}{|c||c|c|c|}
       \hline
        & $m=1$ & $m=2$ & $m=3$ \\ \hline \hline
       $i=1$ & 0 & 0.3346 & 0.8024 \\ \hline
       $i=2$ & 0 & 0.2220 & 0.0659 \\ \hline
       $i=3$ & 0 & 0.2216 & 0.0658 \\ \hline
       $i=4$ & 0 & 0.2219 & 0.0660 \\ 
       \hline
\end{tabular}}\hspace{1cm}
\subtable[$Q_{th} = 10$]
{\begin{tabular}{|c||c|c|c|}
      \hline
        & $m=1$ & $m=2$ & $m=3$ \\ \hline \hline
       $i=1$ & 0 & 0.2828 & 0.9956 \\ \hline
       $i=2$ & 0 & 0.2390 & 0.0004 \\ \hline
       $i=3$ & 0 & 0.2391 & 0.0018 \\ \hline
       $i=4$ & 0 & 0.2391 & 0.0022 \\ 
       \hline
\end{tabular}}
\caption{$\phi_m^i$ of Our Scheduler}
\end{table}

\begin{figure}
   \begin{center}
    \includegraphics[width=0.6\columnwidth]{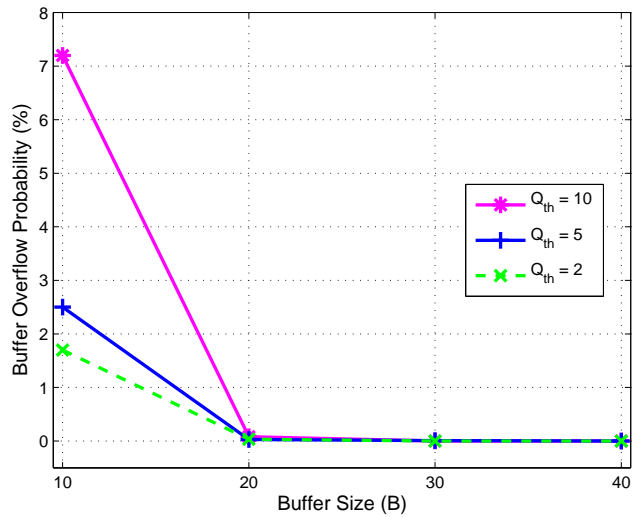}
    \caption{Buffer overflow probability comparison for different values of $Q_{th}$ achieved by our scheduler.}
    \label{fig:our-OF-Prob}
  \end{center}
\end{figure}

Since the LDP only holds for the large value of $B$, for moderately large value $40$ of $B$, we have run our scheduling algorithm under the setting mentioned above, and compared the resultant decay rate of the queue overflow probability with the optimal one ($I_{opt}$) in Fig.~\ref{fig:our-DR}. In this simulation, the values of $\rho_1$ and $\rho_2$ are set as $0$. In the analytical study, we have observed that the decay rate is different for different values of $Q_{th}$, which is validated in this figure. Furthermore, we see that the smaller the value of $Q_{th}$, the better the value of decay rate. However, the decay rate is the best at $Q_{th} = 2$. Now, we want to see the reason behind the relationship between the value of $Q_{th}$ and the decay rate. For this, Table III(a) and Table III(b) show the resultant user-channel state distribution matrix for $Q^{th} = 2$ and $Q^{th}=10$, respectively. To compare our results with the optimal decay rate ($I_{opt}$), we have shown the corresponding user-channel state distribution matrix in Table II. In the analytical study, we argued that the smaller the value of $Q_{th}$, the scheduling algorithm gives more priority to the queue length. This is obvious when we look at Table III, which is at channel state $m = 2$, $\phi_m^1$ is larger for $Q_{th}=2$ comparing with $Q_{th} = 10$. For the similar reason, at channel state $m = 3$, $\phi_m^1$ is smaller for $Q_{th}=2$ comparing with $Q_{th} = 10$. Furthermore, we see in Fig.~\ref{fig:our-DR} that the decay rate for $Q_{th} = 1$ is worse than that with $Q_{th} = 2$, and it implies that only queue aware scheduling rule does not improve the decay rate. We will investigate this issue more in the following discussions. Comparing these tables, it is very obvious that simulation results confer with the analytical study. At $Q_{th}=2$, the resultant user-channel state distribution matrix is more approaching to the optimality comparing with the other cases. Fig.~\ref{fig:our-OF-Prob} shows the buffer overflow probability for different values of $Q_{th}$. The larger the decay rate, the smaller the buffer overflow probability, which is evident in the figure. This figure further proves that the larger the buffer size, the smaller the buffer overflow probability.

\begin{figure}
   \begin{center}
    \includegraphics[width=0.6\columnwidth]{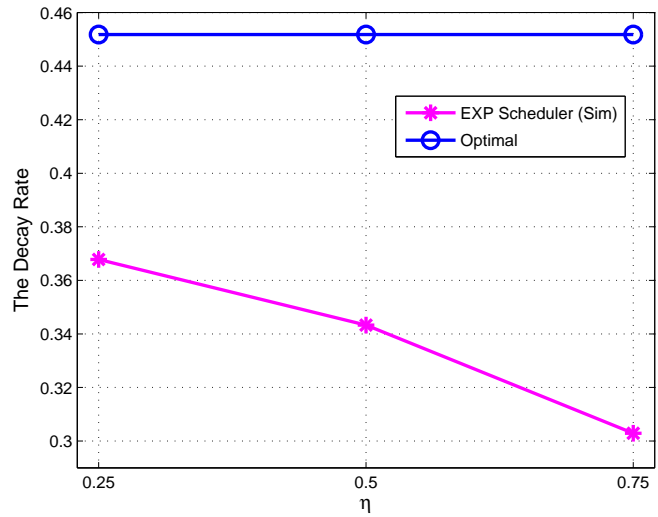}
    \caption{Decay rate comparison between the EXP scheduler ($B = 40$) and the optimal one.}
    \label{fig:EXP-DR}
  \end{center}
\end{figure}

\begin{table}
\centering
\subtable[$\eta = 0.25$]
{\begin{tabular}{|c||c|c|c|}
       \hline
        & $m=1$ & $m=2$ & $m=3$ \\ \hline \hline
       $i=1$ & 0 & 0.3198 & 0.8698 \\ \hline
       $i=2$ & 0 & 0.2269 & 0 \\ \hline
       $i=3$ & 0 & 0.2265 & 0.0614 \\ \hline
       $i=4$ & 0 & 0.2269 & 0.0688 \\
       \hline
\end{tabular}}\hspace{1cm}
\subtable[$\eta = 0.75$]
{\begin{tabular}{|c||c|c|c|}
      \hline
       & $m=1$ & $m=2$ & $m=3$ \\ \hline \hline
       $i=1$ & 0 & 0.2888 & 0.9866 \\ \hline
       $i=2$ & 0 & 0.2371 & 0 \\ \hline
       $i=3$ & 0 & 0.2371 & 0.0066 \\ \hline
       $i=4$ & 0 & 0.2371 & 0.0068 \\ 
       \hline
\end{tabular}}
\caption{$\phi_m^i$ of EXP Scheduler}
\end{table}

\begin{figure}
   \begin{center}
    \includegraphics[width=0.6\columnwidth]{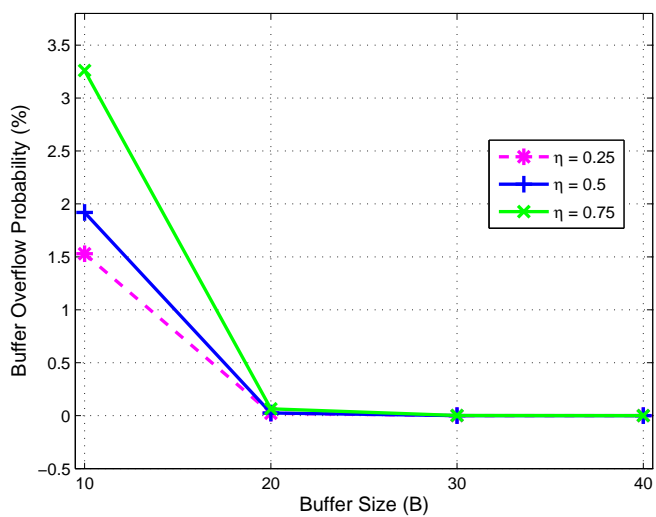}
    \caption{Buffer overflow probability comparison for different values of $\eta$ achieved by the EXP scheduler.}
    \label{fig:EXP-OF-Prob}
  \end{center}
\end{figure}

We also have simulated EXP~\cite{Stolyar08} and MW~\cite{BilalSadiq11} rules for the above setup. At time instant $t$, if the channel state is $m$, EXP rule chooses to serve user $i^*$ following the formula

\[
i^* = \displaystyle\argmax_{i}\mbox{exp}\left[\frac{Q_i(t)}{1 + (\frac{1}{N}\sum_{k}Q_k(t))^\eta}\right]F_m^i,
\]

\noindent
where $\eta$ is a constant parameter that is taken from the interval $(0, 1)$. Similar to our case, we plot the decay rate of the queue overflow probability achieved by this scheduler for different values of $\eta$ in Fig.~\ref{fig:EXP-DR}. Moreover, we present the user-channel state distribution matrix for $\eta = 0.25$ and $\eta = 0.75$ in Table IV(a) and Table IV(b), respectively. We also plot buffer overflow probability for different values of $\eta$ in Fig.~\ref{fig:EXP-OF-Prob}. According to the analytical results of~\cite{Stolyar08}, the optimal value of decay rate for the largest queue overflow probability is independent of $\eta$. However, in the tables, we notice, for larger value of $\eta$, the scheduling algorithm gives more priority to the instantaneous channel rate comparing with the instantaneous queue length, and hence the resultant largest queue length slightly increases for this case as well as the corresponding decay rate decreases. However, from our observation in the simulation, the resultant decay rate does not vary that much for different values of $\eta$. Moreover, the decay rate for $\eta =0.25$ is more approaching to the optimality. This observation is more evident in Table IV(a) and Table IV(b).


\begin{figure}
   \begin{center}
    \includegraphics[width=0.6\columnwidth]{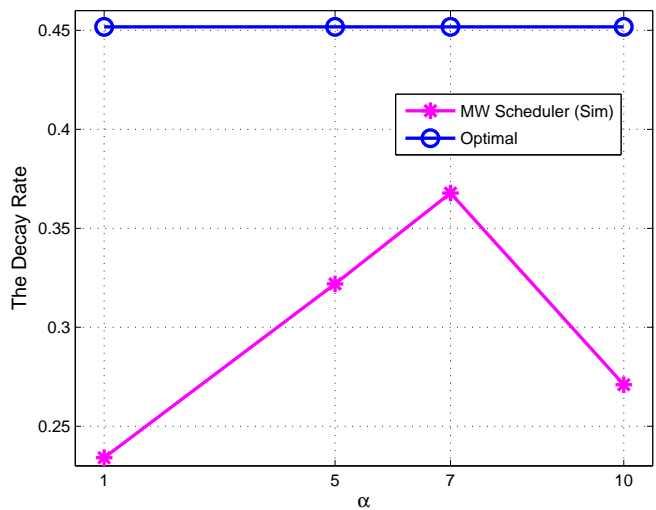}
    \caption{Decay rate comparison between the MW scheduler ($B = 40$) and the optimal one.}
    \label{fig:MW-DR}
  \end{center}
\end{figure}

\begin{table}
\centering
\subtable[$\alpha = 1$]
{\begin{tabular}{|c||c|c|c|}
       \hline
        & $m=1$ & $m=2$ & $m=3$ \\ \hline \hline
       $i=1$ & 0 & 0.2791 & 0.9983 \\ \hline
       $i=2$ & 0 & 0.2403 & 0 \\ \hline
       $i=3$ & 0 & 0.2403 & 0.0009 \\ \hline
       $i=4$ & 0 & 0.2403 & 0.0007 \\ 
       \hline
\end{tabular}}\hspace{1cm}
\subtable[$\alpha = 7$]
{\begin{tabular}{|c||c|c|c|}
      \hline
        & $m=1$ & $m=2$ & $m=3$ \\ \hline \hline
       $i=1$ & 0 & 0.3322 & 0.8146 \\ \hline
       $i=2$ & 0 & 0.2228 & 0 \\ \hline
       $i=3$ & 0 & 0.2224 & 0.0811 \\ \hline
       $i=4$ & 0 & 0.2227 & 0.1042 \\ 
       \hline
\end{tabular}}
\caption{$\phi_m^i$ of MW Scheduler}
\end{table}

\begin{figure}
   \begin{center}
    \includegraphics[width=0.6\columnwidth]{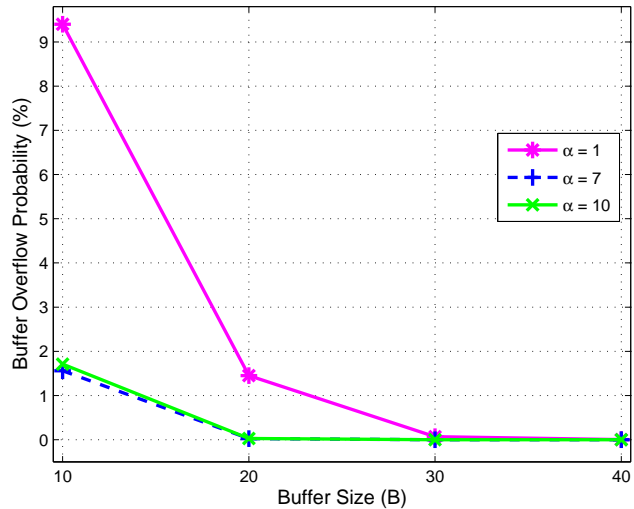}
    \caption{Buffer overflow probability comparison for different values of $\alpha$ achieved by the MW scheduler.}
    \label{fig:MW-OF-Prob}
  \end{center}
\end{figure}

On the other hand, MW rule selects user $i^*$ such that

\[
i^* = \argmax_{i}Q_i(t)^{\alpha}F_m^i.
\]

Here, $\alpha$ is also a constant parameter that ranges from $[1, \infty]$. For this rule, similar to the above two cases, we plot the decay rate and buffer overflow probability in Fig.~\ref{fig:MW-DR} and Fig.~\ref{fig:MW-OF-Prob}, respectively. Table V(a) and Table V(b) present the user-channel state distribution matrix for $\alpha = 1$ and $\alpha = 7$, respectively. Unlike the EXP rule, with this rule, larger value of $\alpha$ makes the scheduling algorithm to be more dependent on the instantaneous queue length comparing with the instantaneous channel rate. However, as we mentioned in the previous paragraphs, the scheduling algorithm that is solely dependent on queue length is not throughput optimal, cannot provide optimal decay rate. To achieve the optimality, one needs to choose the proper value of $\alpha$ that provides the proper balance between the instantaneous queue length and the instantaneous channel rate dependence of the scheduling algorithm. Consequently, the optimal decay rate is achieved at $\alpha = 7$. Table V(a) and Table V(b) further emphasizes this observation. In Fig.~\ref{fig:MW-OF-Prob}, it is further evident that the better the decay rate, the better the buffer overflow probability.


The above simulation results discussed raise some issues on the applicability of the large deviation principle as the probability of any event is an exponential function of its decay rate. From our analytical study and also from~\cite{BilalSadiq11, Stolyar08}, buffer size that is close to $\infty$ fits the characteristics of LDP. This study does not provide much information on what buffer level is enough for the asymptotic system behavior to be accurate. Furthermore, the LDP study only specifies the decay rate of the probability of any event. If the factor in front of the exponential term is unknown, the exact value of the probability from the LDP study is not possible. Hence, one needs to be careful while comparing the performance predicted by a LDP with the actual performance (e.g., buffer overflow probability) of any scheduling algorithm.  

\begin{figure}
   \begin{center}
    \includegraphics[width=0.6\columnwidth]{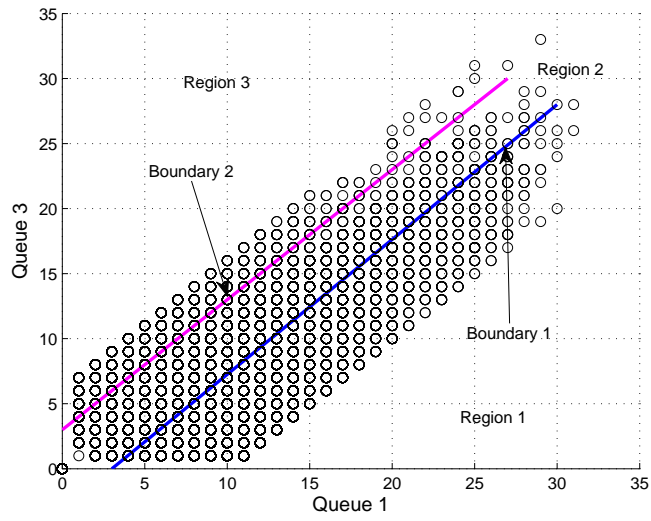}
    \caption{The state space achieved by our scheduler for $Q_{th} = 2$.}
    \label{fig:our-dec-reg-Qth-2}
  \end{center}
\end{figure}

\begin{figure}
   \begin{center}
    \includegraphics[width=0.6\columnwidth]{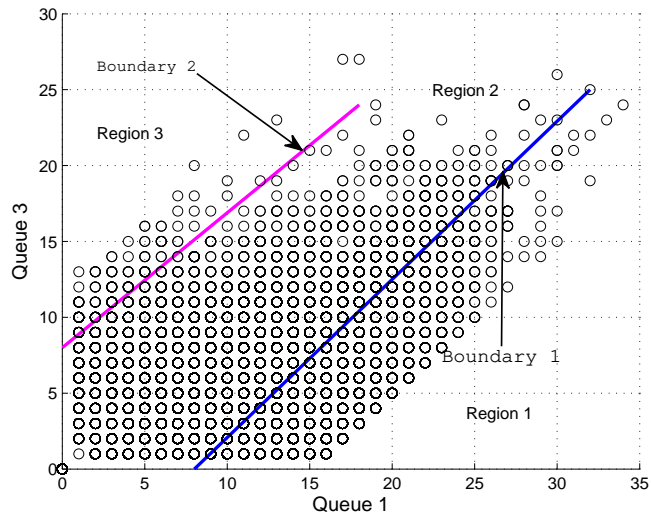}
    \caption{The state space achieved by our scheduler for $Q_{th} = 10$}
    \label{fig:our-dec-reg-Qth-10}
  \end{center}
\end{figure}

To better understand the results, we plot the state space of the simulation (our scheduler) in Fig.~\ref{fig:our-dec-reg-Qth-2} and Fig.~\ref{fig:our-dec-reg-Qth-10} for $Q_{th} = 2$ and $Q_{th} = 10$, respectively. We project the length of two chosen queues $Q_1$ and $Q_3$ on the X-axis and Y-axis, respectively. Based on the scheduling decision, the state space is divided into decision regions. In Region $1$, user $1$ is scheduled based on its queue length irrespective of the channel state. In the similar manner, no matter the channel state, in Region $3$, $Q_3$ is served. In Region $2$, either $Q_1$ or $Q_3$ is served based on their channel state and queue length. Boundary of the decision regions is determined by the scheduling policy. The tiny circles in the figures are the states that have been visited by the system during the simulation.

Region $1$ and Region $3$ are referred to as max-queue regions as in these regions, scheduling decisions are only made based on the queue length. Whereas, Region $2$ is named as the max-rate region as the user instantaneous rate is considered while making the scheduling decision in this region. For our scheduling algorithm, as we decrease $Q_{th}$, the boundaries between the decision regions tend to move to the diagonal line. This trend has two implications. First, as the decision boundaries approach the diagonal line, the algorithm places more emphasis on reducing the largest queue. Apparently, with this setup, the decay rate of the queue overflow probability should be improved significantly. However, the second effect of decreasing $Q_{th}$ is the reduction of the max-rate region. As a result, for smaller value of queue length, it is less likely that the system falls into the max-rate region. With smaller value of $Q_{th}$, the algorithm is unlikely to take the advantage of the increased capacity at small queue lengths which leads to the tendency for the queues to grow. The area of the max-rate region introduces the tradeoff in increasing/decreasing the largest queue length. For this reason, as observed in Fig.~\ref{fig:our-DR}, the decay rate of the overflow probability for $Q_{th} = 1$ is smaller than that with $Q_{th} = 2$. 

\begin{figure}
   \begin{center}
    \includegraphics[width=0.6\columnwidth]{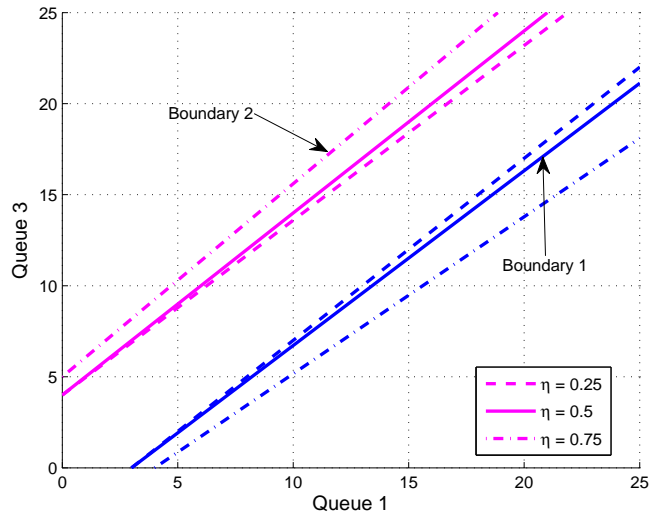}
    \caption{The decision boundaries achieved by the EXP scheduler for various values of $\eta$.}
    \label{fig:EXP-dec-region}
  \end{center}
\end{figure}

Similar to the state space plots achieved from the simulation of our scheduling algorithm, we show the decision boundaries of the state space derived from the simulation of the EXP scheduler in Fig.~\ref{fig:EXP-dec-region} for different values of $\eta$. The slight variation of the decay rate for different values of $\eta$ can be understood when we see different boundary lines. From this figure and the plot in Fig.~\ref{fig:our-DR}, this is evident that with larger values of $\eta$, the larger max-rate region may not be useful in reducing the largest queue length.

\begin{figure}
   \begin{center}
    \includegraphics[width=0.6\columnwidth]{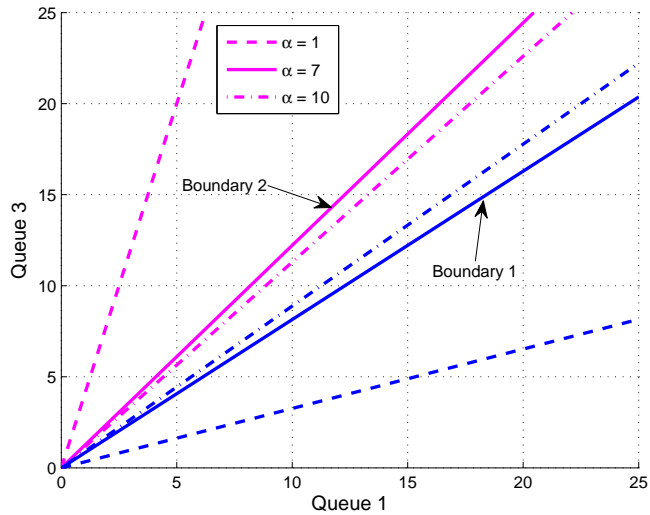}
    \caption{The decision boundaries achieved by the MW scheduler for various values of $\alpha$.}
    \label{fig:MW-dec-region}
  \end{center}
\end{figure}

For different values of $\alpha$, we also plot the decision boundaries of the simulated state space obtained by the MW scheduling algorithm. We argued previously that the MW rule provides more emphasis on the largest queue length for larger values of $\alpha$, and fails to take the advantage of the instantaneous channel rate. This observation is further proved in this figure. However, with smaller values of $\alpha$, the algorithm exaggeratedly gives more emphasis on the instantaneous use rate, and hence the largest queue length increases consequently. As a result, the best decay rate of the overflow probability is obtained at $\alpha = 7$.

\section{Conclusion}
\label{sec:concl}

In this paper, we have analyzed the performance of the scheduling algorithm proposed in~\cite{RRuby15}. Because of the complex coupling between input and output metrics of a scheduling algorithm, it is difficult to analyze such algorithm using conventional methods. Since the scheduling algorithm is specifically designed for QoS based traffic, we have mapped the probability that the scheduling algorithm does not meet the QoS bound to the queue overflow probability. Then, we have used LDP to determine the bound of this probability. We have proved that for certain value of the parameter $Q_{th}$, the scheduling algorithm converges to the optimal possible algorithm which can reduce the queue overflow probability to the smallest possible level. Finally, through the numerical results, we have proved our analytical study comparing with other scheduling algorithms.

%






\end{document}